\documentstyle[preprint,aps,epsfig]{revtex}
\tighten
\newcommand{\beq}{\begin{eqnarray}}
\newcommand{\eeq}{\end{eqnarray}}

\newcommand{\bsz}{b\overline{s}Z}
\newcommand{\bsdnn}{B \to X_{s,d} \; \nu \overline{\nu}}
\newcommand{\bsnn}{B \to X_{s} \; \nu \overline{\nu}}
\newcommand{\bdnn}{B \to X_{d}\; \nu \overline{\nu}}
\newcommand{\calb}{{\cal B}}

\newcommand{\mhp}{m_{H^+}}
\newcommand{\ssa}{\sin^2\theta_W}
\newcommand{\brbsnn}{{\cal B}(B \to  X_{s}\nu \bar{\nu})}

\title{Rare decay $B \to X_{s}\; \nu \bar \nu$ in the two-Higgs-doublet model of type-III }
\author{ Zhenjun Xiao $^1$
\thanks{Email address: zjxiao@email.njnu.edu.cn}, Liping Yao$^{2}$ \\
{\small $^1$ Department of Physics, Nanjing Normal University,
Nanjing, Jiangsu 210097, P.R.China} \\
{\small $^2$ Department of Physics, Henan Normal University, Xinxiang, Henan 453002, P.R.China} }
\date{\today}
\begin{document}
\maketitle
\begin{abstract}
In this paper, we calculated the new physics contribution to theoretically very clean
rare decay $B\to X_{s} \nu \bar{\nu}$ in the general two-Higgs-doublet model (model III).
Within the considered parameter space, we found that
(a) the new physics contribution can provide one to two orders of enhancement  to the
branching ratio $\calb(B\to X_s \nu \bar{\nu})$ and can saturate the experimental bound
on $\calb(B \to X_s \nu \bar{\nu})$ in some regions of the parameter space;
(b) besides the CLEO data of $B \to X_s \gamma$, the ALEPH upper
limit on $\calb (B \to X_s \nu \bar{\nu})$ also lead to further constraint
on the size of the Yukawa coupling $\lambda_{tt}$: $\lambda_{tt}< 6.4$ for
$\lambda_{bb}=2.7$ and $\mhp=200$ GeV.
\end{abstract}

\vspace{0.5cm}



\newpage

In the standard model (SM), the rare decays $\bsdnn$ are fully dominated by the
$Z^0$-penguin and box diagrams involving top quark exchanges. The charm quark contribution
and the long distance contributions are negligible, and the theoretical uncertainties related the
renormalization scale dependence of running quark mass can be essentially neglected
after the inclusion of next-to-leading order corrections\cite{buras01}.
These decays are theoretically very clean processes in the field of rare B-decays and are also
sensitive to the new physics beyond the SM\cite{he88,grossman96,huang01}.

The decays $\bsdnn$ have been thoroughly studied in \cite{grossman96} and reviewed recently
in \cite{buras01}. Normalizing to the semi-leptonic
branching ratio ${\cal B}(B \to X_c e \bar{\nu})$ and summing over the three neutrino flavors,
one finds\cite{buras01}
\beq
{\cal B}(B \to X_q \nu \bar{\nu}) = {\cal B}(B \to X_c e \bar{\nu})
 \frac{3 \alpha_{em}^2}
{4 \pi^2\sin^4\theta_{W}} \frac{\mid V_{tq} \mid ^2}{\mid V_{cb} \mid ^2}
\frac{|X(x_{t})|^2}{f(z)} \frac{\overline{\eta}}{\kappa(z)}
\label{brs1}
\eeq
where $q=(d,s)$, $x_t=m_t^2/m_W^2$ ($m_t$ and $m_W$ are the masses of the top quark and
$W$ gauge boson, respectively),  $V_{ij}$ are the elements of
the Cabibbo-Kabayashi-Maskawa (CKM) mixing matrix.  $f(z)$ and $\kappa(z)$
($\bar{\eta}=\kappa(0)$) with $z=m_c^{pole}/m_b^{pole}$ are the phase-space and quantum
chromodynamics(QCD) correction factors for the decay $B \to X_{c} e \overline{\nu}$
\cite{buras96}. For $z=0.29$, one finds $f(z)=0.54$ and $\kappa(z)=0.88$.
The function $X(x_t)$ describes the top quark contribution and is given
by \cite{buras96}
\beq
X(x_t)&=&X_0(x_t) + \frac{\alpha_s}{4\pi} X_1(x_t)\label{eq:xxt}
\eeq
with
\beq
X_0(x)=-\frac{x}{8}\left [ \frac{2+x}{1-x} + \frac{6 - 3 x}{(1-x)^2}\ln[x]\right ],
\eeq
and the QCD correction $X_1(x_t)$ can be found in \cite{buras96}. As mentioned
previously, the SM predictions for $\bsdnn$ are remarkably free from
uncertainties. All the parameters entering in Eq.(\ref{brs1}) are known with
good accuracy. Since the decay $\bdnn$ is further suppressed by the ratio
$|V_{td}/V_{ts}|^2 \sim 0.04$, the decay $\bsnn$ is more interesting experimentally.
We here consider the decay $\bsnn$ only. The experimental upper limit
\beq
{\cal B}(B \to X_s \nu \bar{\nu}) < 6.4 \times 10^{-4} \ \ at \ \ 90\% C.L.\label{eq:snnexp}
\eeq
has been reported last year by ALEPH collaboration\cite{aleph01} at LEP, which
is close to the SM prediction of $(3.5 \pm 0.7 )\times 10^{-5}$\cite{buras01}.
The decay $\bsnn$  may be accessible at B factories if background problems
can be resolved. The decay $\bsnn$ can probe many new physics scenarios and has been
studied for example in the Technicolor models \cite{xiao99}, the two-Higgs-doublet models (2HDM)
of type-II and the supersymmetric models \cite{buras01b}.

In this paper, we calculate the new physics contributions to the rare decay
$\bsnn$ due to the effective $\bsz$ coupling induced by
the charged-Higgs penguin diagrams in the so-called model III: the
2HDM with flavor changing (FC) couplings\cite{atwood97}.
In the 2HDM, the tree level flavor changing
scalar currents are absent if one introduces an {\it ad
hoc} discrete symmetry to constrain the 2HDM scalar potential and
Yukawa Lagrangian. Lets consider a Yukawa Lagrangian
of the form\cite{atwood97}
\beq
{\cal L}_Y &=&
\eta^U_{ij}\bar{Q}_{i,L} \tilde{\phi_1}U_{j,R} +
\eta^D_{ij}\bar{Q}_{i,L} \phi_1 D_{j,R}
+\xi^U_{ij}\bar{Q}_{i,L} \tilde{\phi_2}U_{j,R}
+\xi^D_{ij}\bar{Q}_{i,L} \phi_2 D_{j,R}+ h.c., \label{leff}
\eeq
where $\phi_{i}$ ($i=1,2$) are the two Higgs doublets,
 $\tilde{\phi}_{1,2}=i\tau_2 \phi^*_{1,2}$, $Q_{i,L}$  with
 $i=(1,2,3)$ are the left-handed quarks,
$U_{j,R}$ and $D_{j,R}$  are the right-handed  up- and down-type quarks,
while $\eta^{U,D}_{i,j}$  and $\xi^{U,D}_{i,j}$ ($i,j=1,2,3$ are
family index ) are generally the nondiagonal matrices of the  Yukawa
coupling. By imposing the discrete symmetry $(\phi_1 \to - \phi_1,
\phi_2 \to \phi_2, D_i \to - D_i, U_i \to  \mp U_i) $ one obtains
the so called model I and II. One finds the model III if no discrete
symmetry is imposed. In model III, there are also five physical
Higgs bosons: the charged
scalar $H^{\pm}$, the neutral CP even scalars $H^0$ and $h^0$
and the CP odd pseudoscalar $A^0$.

After the rotation that
diagonalizes the mass matrix of the quark fields, the Yukawa
Lagrangian of quarks  are the form \cite{atwood97},
\beq
{\cal L}_Y^{III} =
\eta^U_{ij}\bar{Q}_{i,L} \tilde{\phi_1}U_{j,R} +
\eta^D_{ij}\bar{Q}_{i,L} \phi_1 D_{j,R}
+\hat{\xi}^U_{ij}\bar{Q}_{i,L} \tilde{\phi_2}U_{j,R}
+\hat{\xi}^D_{ij}\bar{Q}_{i,L} \phi_2 D_{j,R} + h.c.,
\label{lag3}
\eeq
where $\eta^{U,D}_{ij}=m_i\delta_{ij}/v$ correspond to the diagonal
mass matrices of quarks and $v\approx 246 GeV$ is the vacuum
expectation value of $\phi_1$, while the neutral and charged FC
couplings will be \cite{atwood97}
\beq
\hat{\xi}^{U,D}_{neutral}&=& \xi^{U,D}, \ \ \hat{\xi}^{U}_{charged}= \xi^{U}V, \ \
\hat{\xi}^{D}_{charged}= V \xi^{D}, \label{cxiud}
\eeq
where $V_{CKM}$ is the Cabibbo-Kabayashi-Maskawa mixing matrix, and
\beq
\xi^{U,D}_{ij}=\frac{\sqrt{m_im_j}}{v} \lambda_{ij}.\label{xiij}
\eeq
where the coupling parameters $\lambda_{ij}$ ($i,j=(1,2,3)$ are the generation index)
are free parameters to be determined by experiments.
As pointed in Ref.\cite{atwood97}, the data of $K^0-\bar{K}^0$ and $B_d^0-\bar{B}_d^0$ mixing
processes put severe constraint on the FC couplings involving the first two generations of
quarks. One therefore assume that:
\beq
\lambda_{ij}=0,  \ \ for \ \ i=1,2, \ \  and \ \ j=1,2,3.
\eeq

From the CERN $e^+ e^-$ collider (LEP) and Tevatron searches for charged Higgs bosons
\cite{lep2}, the combined constraint in the $(\mhp, \tan{\beta})$ plane has been
given in Ref.\cite{pdg00}: the direct lower limit is $\mhp > 77$ GeV.
In this paper, we consider Chao, Cheung and Keung (CCK)
scenario of the model III\cite{chao99}: only the couplings
$\lambda_{tt}$ and $\lambda_{bb}$ are non-zero. In this scenario, the existence of a
charged Higgs boson with $\mhp \sim 200$ GeV is still allowed \cite{chao99,xiao2000,prd207}.

\begin{figure}[t]
\vspace{-40pt}
\begin{minipage}[]{\textwidth}
\centerline{\epsfxsize=0.9\textwidth \epsffile{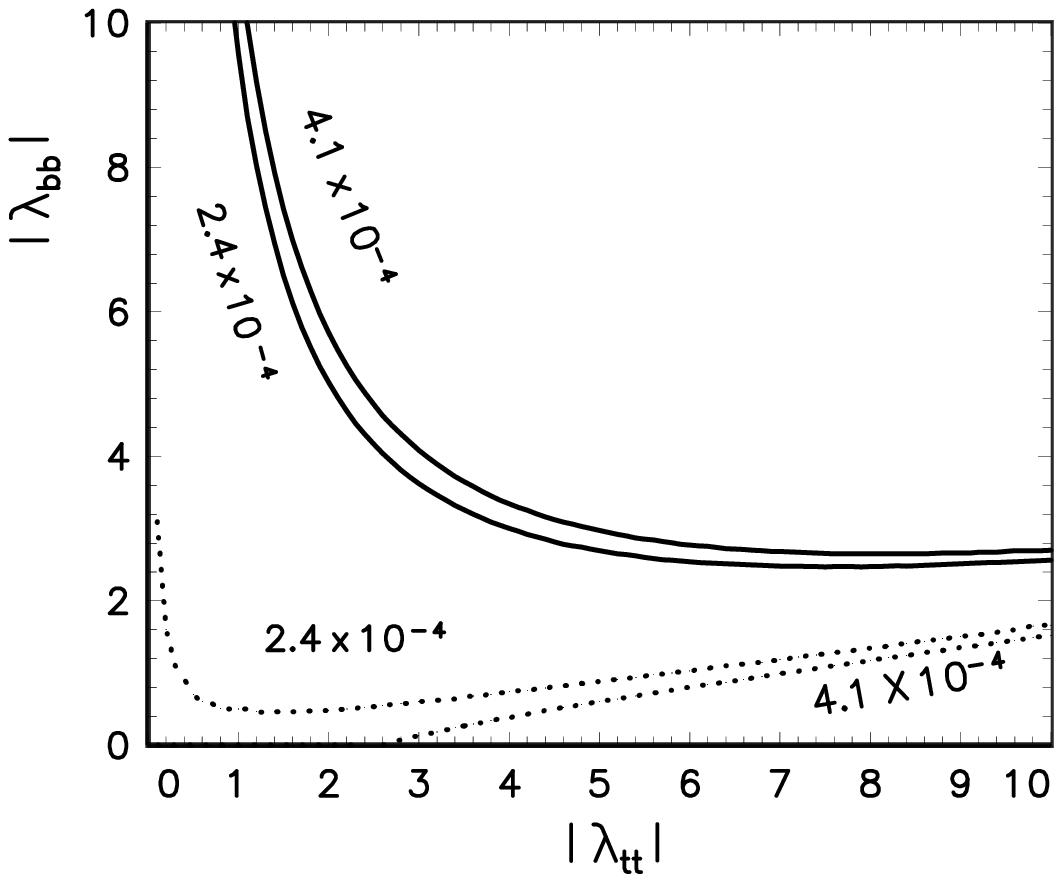}}
\vspace{-40pt}
\caption{Contour plot of the branching ratio $\calb ( B \to X_s \gamma)$ versus $\lambda_{tt}$
and $\lambda_{bb}$ for $\mhp =200$ GeV and $\theta=0^\circ$.
The narrow regions between two dotted curves and two solid
curves are still allowed by the data: $2.4\times 10^{-4} \leq  \calb (B \to X_s \gamma)
\leq 4.1\times 10^{-4}$.}
\label{fig:fig1}
\end{minipage}
\end{figure}

From the CLEO data of $\calb (B \to X_s \gamma)$, some constraint on $\mhp$ in model
III can also be derived \cite{chao99}. New precision measurements
of $B \to X_s \gamma$ are reported recently by CLEO \cite{cleo01} and by Belle
Collaboration \cite{belle01}. Combining with previous determinations\cite{pdg00}, the world
average as given in Ref.\cite{isidori01} reads: $\calb (B \to X_s \gamma)^{exp}=(3.23 \pm 0.42)
\times 10^{-4}$. At the $2\sigma$ level, we have $2.4\times 10^{-4} \leq
\calb (B \to X_s \gamma) \leq 4.1 \times 10^{-4}$.
Fig.\ref{fig:fig1} is the contour plot of the branching ratio $\calb ( B \to X_s \gamma)$
versus $|\lambda_{tt}|$ and $|\lambda_{bb}|$ for $\mhp =200$ GeV and
$\theta=\theta_{tt}-\theta_{bb}=0^\circ$.
It is easy to see that most part of the parameter space of $|\lambda_{tt}|$ and $|\lambda_{bb}|$
has been excluded, only the narrow regions between two dots curves and two solid
curves are still allowed by the data of $B \to X_s \gamma$.
In the region of $\lambda_{tt} \geq 5$,
the upper (lower) band has a weak (moderate) dependence on the value of $\lambda_{bb}$:
$\lambda_{bb} = 2.7\pm 0.2$ for the upper band
and $\lambda_{bb}=1.0^{+0.6}_{-0.4}$ for the lower band.

Under the assumption of neglecting all non-Z-mediated contributions, Buchalla
et al. \cite{buchalla01} studied the constraints on the effective couplings $Z_{bs}^{L,R}$
obtained by comparing the theoretical predictions of the branching ratios of
$B \to X_s \nu \bar{\nu}, X_s l^+ l^-$ and $B \to K^* \mu^+ \mu^-$ decay modes with the
experimental measurements\cite{aleph01,cleo98,cdf99}\footnote{For $B \to X_s \nu \bar{\nu}$
decay, we here use the new ALEPH measurement \cite{aleph01} $\calb (B \to X_s \nu \bar{\nu}
)< 6.4 \times 10^{-4}$ instead of the old one as used in Ref.\cite{buchalla01}.}:
\beq
 \sqrt{|Z_{bs}^L|^2 + |Z_{bs}^R|^2} \lesssim 0.22, && \ \ \  from \ \ \  \calb (B \to X_s \nu \bar{\nu}
)< 6.4 \times 10^{-4}, \\
 \sqrt{|Z_{bs}^L|^2 + |Z_{bs}^R|^2} \lesssim 0.15,  && \ \ \ from \ \ \ \calb (B \to X_s l^+ l^-)
< 4.2 \times 10^{-5}, \\
 |Z_{bs}^{L,R}| \lesssim 0.13, && \ \ \  from \ \ \ \calb (B \to K^* \mu^+ \mu^-
)^{n.r.} < 4.0 \times 10^{-6}
\eeq
Among the three modes, $B \to X_s \nu \bar{\nu}$ decay is the cleanest mode.
Since the SM contributions are $Z_{bs}^{L}=V_{tb}^* V_{ts} C_0(x_t) \sim 0.04$
and $Z_{bs}^{R}=0$, a new physics
enhancement to the branching ratio $\calb (B \to X_s \nu \bar{\nu})$ as large as a factor
of $(0.22/0.04)^2 \sim 30$ is still allowed by the first bound.
The third bound is stronger but subject to large theoretical uncertainties in the form factors
and the assumptions on the non-perturbative non-resonant contributions, and therefore less reliable.
The detailed discussions for the new physics enhancement to  $\calb(B \to X_s \nu \bar{\nu})$
and the possible constraints on the FC couplings of model III will be given below.

\begin{figure}[t]
\vspace{-40pt}
\begin{minipage}[]{\textwidth}
\centerline{\epsfxsize=1.2\textwidth \epsffile{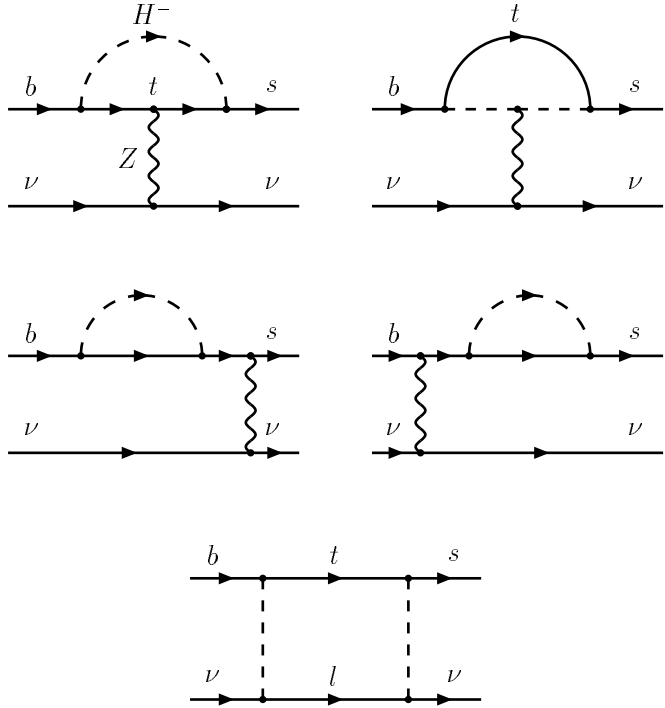}}
\vspace{-200pt}
\caption{The typical new $Z^0$-penguin, self-energy  and box diagrams for
$B \to X_s \nu \bar{\nu}$ decay in model III.}
\label{fig:fig2}
\end{minipage}
\end{figure}

For the decay processes considered here, as illustrated in Fig.\ref{fig:fig2},
the new $Z$-penguin diagrams with internal top quarks obtained by replacing the $W$ gauge
boson with the charged-Higgs boson dominate the new physics corrections. The
charged-Higgs box diagram does not contribute because the Yukawa coupling $H^+l\nu$ is zero.
The neutral Higgs bosons also do not contribute at tree level or one-loop level in the CCK
scenario of model III \cite{prd207}.
In the calculation, we will use dimensional regularization to regulate
all the ultraviolet divergences in the virtual loop corrections
and adopt the modified minimal subtracted ($\overline{MS}$)
renormalization scheme.

By analytical evaluations of the Feynman diagrams, we find the effective $\bsz$
vertex induced by the charged-Higgs exchanges,
\beq
\Gamma_{Z_{\mu}} =
\frac{1}{16 \pi^2}\frac{g^3}{\cos\theta_W}\, V_{ts}^*V_{tb}\,
\overline{s_L}\, \gamma_{\mu}\, b_L\; C_0^N(y_t) \label{bsza}
\eeq
with
\beq
C_0^N(y_t) &=& -\frac{x_t}{16}\left \{
\left [ \frac{y_t}{1-y_t} + \frac{y_t }{(1-y_t)^2} \ln[y_t]  \right ] |\lambda_{tt}|^2\right.
\nonumber\\
&& \left. + \frac{4m_b^2 \ssa}{3m_t^2}
\left [ \frac{3y_t -y_t^2}{4 (1-y_t)^2} + \frac{y_t}{(1-y_t)^3}\ln[y_t]
\right ] |\lambda_{tt}|^2 \right.  \nonumber \\
&& \left. +  \frac{m_b^2}{m_t^2} \left [ \frac{3y_t -y_t^2}{4 (1-y_t)^2}
+ \frac{y_t}{(1-y_t)^3}\ln[y_t]\right. \right.  \nonumber \\
&& \left. \left.
+ \left (  1-\frac{4}{3}\ssa \right )
\left ( \frac{y_t}{1-y_t} + \frac{y_t }{(1-y_t)^2} \ln[y_t]   \right )
\right ]   |\lambda_{tt}\lambda_{bb}|e^{i\theta}  \right \} \label{eq:cza}
\eeq
where $y_t={m_t}^2/{\mhp}^2$, $\theta_W$ is the Weinberg angle, and
$\theta=\theta_{tt}-\theta_{bb}$. It is easy to see that the value of $C_0^N$
( i.e, the size of new physics corrections) is controlled by the first term of
Eq.(\ref{eq:cza}). The second and third terms of Eq.(\ref{eq:cza}) are strongly
suppressed by the factor of $m_b^2/m_t^2 \approx 8 \times 10^{-4}$. In other words, the parameter
$\lambda_{tt}$ dominates the new physics contribution, while $\lambda_{bb}$ and
$\theta$ play a minor role only.
When the new physics contributions are taken into account, the $X$ function
in Eqs.(\ref{eq:xxt}) takes the form
\beq
X(x_t, y_t) &=& X(x_t) + C_0^N(y_t), \label{eq:xxtb}
\eeq
where function $X(x_t)$ has been given in Eq.(\ref{eq:xxt}).

In the numerical calculations, the following  parameters will be used as the
standard input: $M_W=80.42$GeV, $\alpha_{em}=1/129$, $\ssa =0.23$,
$m_c=1.5$ GeV, $m_b=4.88$ GeV, $m_t \equiv \overline{m_t}(m_t) = 168\pm 5 $ GeV,
$\Lambda^{(5)}_{\overline{MS}}=0.225$GeV, ${\cal B}(B \to X_{c}e \overline{\nu})=10.4\pm 0.6\%$,
$A=0.847$, $\lambda=0.2205$, $R_b=0.38 \pm 0.08$,
$ \gamma=( 60 \pm 20 )^\circ$ and $\theta=0^\circ -30^\circ$.
For the definitions and values of these input parameters, one can see
Refs.\cite{buras01,pdg00}. We have neglected the small errors of
those well measured quantities, but keep the errors for remaining parameters in
order to estimate the uncertainties of the theoretical predictions.
We treat the masses of charged-Higgs boson as semi-free parameters $\mhp=200 \pm 100$ GeV.

\begin{figure}[t]
\vspace{-40pt}
\begin{minipage}[]{\textwidth}
\centerline{\epsfxsize=0.9\textwidth \epsffile{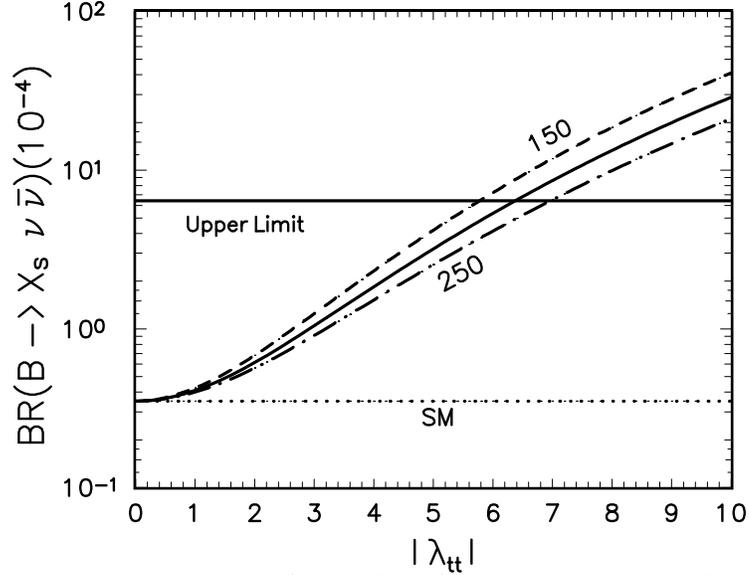}}
\vspace{-40pt}
\caption{Plots of the branching ratio $\calb (B \to X_s \nu \bar{\nu})$ versus
$\lambda_{tt}$ in the SM and model III for $\lambda_{bb}=2.7$,
and $\mhp=150$ (short-dashed curve), $200$
(soild curve) and $250$ GeV (dot-dashed curve), respectively.
The dotted line is the SM prediction, while the upper solid line corresponds to
the ALEPH upper limit: $\calb (B \to X_s \nu \bar{\nu}) < 6.4 \times 10^{-4}$.}
\label{fig:fig3}
\end{minipage}
\end{figure}

Fig.\ref{fig:fig3} shows the $\lambda_{tt}$ dependence of the branching ratio
$\brbsnn$ in the SM and model III for $\lambda_{bb}=2.7$, $\theta=0^\circ$,
and $\mhp=150$ (short-dashed curve), $200$ (solid curve) and $250$ GeV (dot-dashed curve),
respectively. The dotted line in Fig.3 is the SM prediction: $\brbsnn= 3.5 \times
10^{-5}$.
The upper solid line corresponds to the ALEPH upper limit: $\calb (B \to X_s \nu
\bar{\nu}) < 6.4 \times 10^{-4}$. The new physics contribution in model III can provide
one to two orders of enhancement to the branching ratio $\brbsnn$. Furthermore, the constraint on
$\lambda_{tt}$ can be read off directly from Fig.3: $\lambda_{tt} \leq 6.4$ for
$\mhp \approx 200$ GeV, which is complementary to the limits obtained from the
$B \to X_s \gamma$ data.

In order to reduce the effects of uncertainties of input parameters, we
can denote the model III prediction normalized to the SM results by $R$,
\beq
R(B \to X_{s} \nu \bar{\nu}) =\frac{ {\cal B}(B \to X_{s} \nu \bar{\nu})^{III} }{
{\cal B}(B \to X_{s} \nu \bar{\nu})^{SM}} =\frac{|X(x_t,y_t)|^2}{|X(x_t)|^2}
\eeq
The uncertainties of most input parameters are clearly cancelled out in such ratio.

\begin{figure}[t]
\vspace{-40pt}
\begin{minipage}[]{\textwidth}
\centerline{\epsfxsize=0.9\textwidth \epsffile{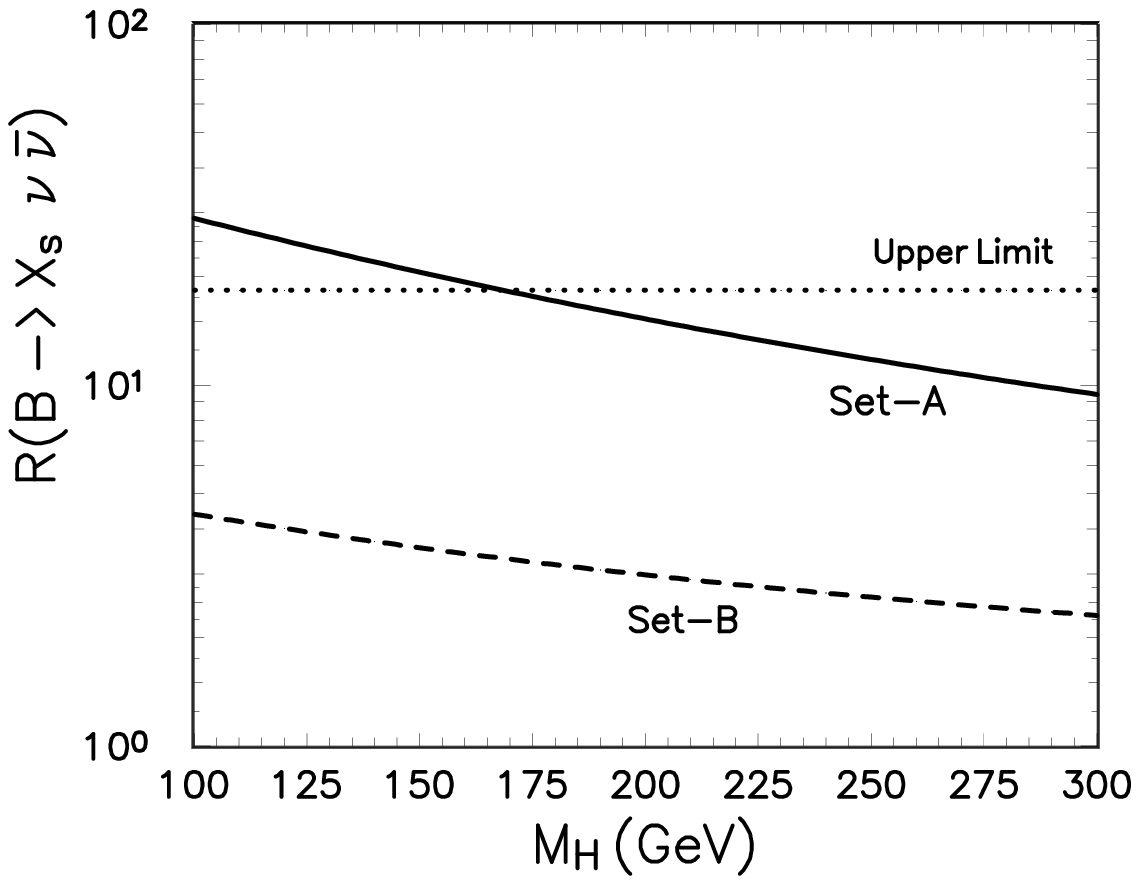}}
\vspace{-40pt} \caption{Plots of the ratio $R(B \to X_s \nu
\bar{\nu})$ versus $\mhp$ for Set-A (solid curve)
and Set-B (short-dashed curve) Yukawa couplings in model III.
The dotted line  refers to the ALEPH upper limit: $\brbsnn < 6.4 \times 10^{-4}$.}
\label{fig:fig4}
\end{minipage}
\end{figure}

In Fig.\ref{fig:fig4}, we show the dependence of the ratio $R$ on the mass $\mhp$ by using
the input parameters as specified before and setting two representative sets of Yukawa couplings
allowed by the CLEO data of $B \to X_s \gamma$:
Set-A: $(\lambda_{tt},\lambda_{bb})=(6,2.7)$(solid curve); and
Set-B: $(\lambda_{tt}, \lambda_{bb})=(3,0.5)$(short-dashed curve). The dotted
line in Fig.4 corresponds to the ALEPH upper limit $R(B \to X_s \nu \bar{\nu})^{exp} <
18.3$.
The ratio R has a very weak dependence on both the ratio $m_t/m_W$ and the angle $\theta$
for $m_t(m_t)=168\pm 5$ GeV and $\theta=0^\circ -30^\circ$.

Although the experimental upper bound on $\brbsnn$ is still a factor of 18 above the SM
expectation, this upper bound has lead to interesting constraints on the parameter space
of model III. The $B \to X_{s} \nu \bar{\nu}$ decay can probe many new physics
scenarios\cite{grossman96}, and deserve the maximum of attention.
We know that the measurement of $\bsnn$ decay is experimentally very
challenging, the gap between SM expectation and experimental limits could decrease in the
next few years at B-factory experiments. The observation of these decays in the near
future will enable us to confirm or exclude the new physics contributions, or at least
put some stringent constraints on two-Higgs doublet models and other new physics models.

In summary, we calculated the new physics contribution to theoretically very clean
rare decay $B\to X_{s} \nu \bar{\nu}$ in the third type of two-Higgs-doublet models.
Within the considered parameter space, we found that:
(a) the new physics contribution can provide one to two orders of enhancements  to the
rare decays $B\to X_{s} \nu \bar{\nu}$ and can saturate the experimental bound
on $\calb(B \to X_s \nu \bar{\nu})$ in some regions of the parameter space;
(b) besides the CLEO data of $B \to X_s \gamma$, the ALEPH upper
limit on $\calb (B \to X_s \nu \bar{\nu})$ also lead to further constraint
on the size of $\lambda_{tt}$: $\lambda_{tt}< 6.4$ for
$\lambda_{bb}=2.7$ and $\mhp=200$ GeV.

\section*{ACKNOWLEDGMENTS}
Xiao Zhenjun acknowledges the support by the National Natural Science Foundation of
China under Grant No.10075013, and by the Research Foundation of Nanjing Normal
University under Grant No.214080A916.

\newpage

\newpage

\newpage

\newpage

\begin{thebibliography}{99}

\bibitem{buras01}
A.J.~Buras, in {\it Flavour Physics: CP violation and rare Decays}, Lectures given at
International School of Subnuclear Physics, Erice, Italy, 2000, hep-ph/0101336.

\bibitem{he88}
X.G.He, T.D. Nguyen, and R.R. Volkas, Phsy.Rev. {\bf D38}(1988)814;
W.Sikba and J.Kalinowski, Nucl.Phys. {\bf B404}, (1993)3.

\bibitem{grossman96}
Y.Grossman, Z.ligeti and E.Nardi, Nucl.Phys. {\bf B465} (1996)369.

\bibitem{huang01}
Y.B.Dai, C.S.Huang and H.W.Huang, Phys.Lett. {\bf B390} (1997)257;
K.S.Babu, C.Kolda, Phys.Rev.Lett. {\bf 84} (2000)228;
C.S.Huang, W.Liao, Q.S.an, and S.H.Zhu, Phys.Rev. {\bf D63} (2001)114021;
C.S.Huang and W. Liao, Phys.Lett. {\bf B525} (2002)107.

\bibitem{buras96}
G.Buchalla, A.J.Buras and M.E. Lautenbacher, Rev.Mod.Phys. {\bf 68} (1996)1125.

\bibitem{aleph01}
R.Barate {\it et al.,} ALEPH Collab., Eur.Phys.J. {\bf C19} (2001)213.

\bibitem{xiao99}
Z.J.Xiao, L.X.L\"u, H.K.Guo and G.R.Lu, Chin.Phys.Lett. {\bf 16} (1999)88;
Z.J.Xiao, L.Q.Jia, L.X.L\"u and G.R.Lu, Communi.Theor.Phys. {\bf 33} (2000)269.

\bibitem{buras01b}
A.J.Buras, P.Gambino, M.Gorbahn, S.J\"ager and L. Silvestrini, Nucl.Phys. {\bf B592} (2001)55;
C.Bobeth, A.Buras, F.Kr\"uger and J.Urban, Nucl.Phys. {\bf B630}(2002)87.

\bibitem{atwood97}
D.Atwood, L.Reina and A.Soni, Phy.Rev. {\bf D55} (1997)3156, and reference therein.

\bibitem{lep2}
L3 Collaboration, M.Acciarri  {\it et al.}, Phys.Lett. {\bf B466} (1999)71;
D0 Collaboration, B.Abbott {\it et al.}, Phys.Rev.Lett. {\bf 82}, (1999)4975;
CDF Collaboration, T.Affolder {\it et al.}, Phys.Rev. {\bf D62} (2000)012004.

\bibitem{pdg00}
Particle Data Group, D.E.Groom {\it et al.}, Eur.Phys.J. {\bf C15} (2000)1.

\bibitem{chao99}
D. B. Chao, K. Cheung and W.Y. Keung, Phys.Rev. {\bf D59} (1999)115006.

\bibitem{xiao2000}
Z.J.Xiao, C.S.Li and K.T. Chao, Phys.Rev. {\bf D62} (2000)094008;
Phys.Lett. {\bf B473} (2000) 148.

\bibitem{prd207}
Z.J.Xiao, C.S.Li and K.T. Chao, Phys.Rev. {\bf D63} (2001)074005.

\bibitem{cleo01}
CLEO Collaboration, S. Chen, {\it et al.}, Phys.Rev.Lett. {\bf 87}
(2001)251807.

\bibitem{belle01}
Belle Collaboration, K.Abe, {\it et al.}, Phys.Lett. {\bf B511} (2001)151.

\bibitem{isidori01}
G.Isidori, Lepton Photon 2001, Rome, Italy, July 2001; hep-ph/0110255.

\bibitem{buchalla01}
G.~Buchalla, G.Hiller and G.Isidori, Phys.Rev. {\bf D63}(2001)014015.

\bibitem{cleo98}
The CLEO Collaboration,  S.~Glenn {\it et al.}, Phys.Rev. Lett. {\bf 80}(1998)2289.

\bibitem{cdf99}
CDF Collaboration,  T.~Affolder {\it et al.}, Phys.Rev. Lett. {\bf 83}(1999)3378.

\end{thebibliography}
\end{document}